\documentclass[11pt]{article}

\usepackage{amsmath}

\begin{document}

\begin{center}
{\bf  E.M. Ovsiyuk\footnote{e.ovsiyuk@mail.ru}, K.V. Kazmerchuk \\[10mm]
NONRELATIVISTIC APPROXIMATION\\FOR QUASI-PLANES WAVES OF A SPIN 1 PARTICLE\\IN LOBACHEVSKY SPACE}

{\small  Mozyr State Pedagogical University named after I.P. Shamyakin, Belarus}

\end{center}

\date{}

\begin{abstract}

Spin 1 particle in Pauli approximation is investigated on the background of the curved space of constant negative curvature, Lobachevsky space. Nonrelativistic approximation is performed in the system of 10 equations resulted from separating the variables in Duffin--Kemmer equation specified in quasi-cartesian coordinates. The problem is solved exactly in Bessel functions, the quantum states are determined by four quantum numbers. The treatment is substantially based on the use of a generalized helicity operator in Lobachevsky space model.
\end{abstract}

Let us start with the system of equations obtained  after
separation of the bariables in Duffin--Kemmer equation for a spin
1 field in quasi-cartesian  coordinates of the Lobachevsky space
\cite{1}
$$
 dS^{2}= dt^{2} - e^{-2z} ( dx^{2} + dy^{2} )
- dz^{2} \; ,\qquad \Psi = e^{-i \epsilon  t } \; e^{i a x} \;
e^{ib y} \left |
\begin{array}{c}   \Phi_{0}(z)  \\ \Phi_{j} (z) \\  E_{j} (z) \\
H_{j} (z)
\end{array} \right |;
$$

\noindent 10 equations are
$$
 \gamma (i a  - b ) e^{z}\,E_{1} -
\gamma (ia +b) e^{z} E_{3} - ({d \over dz} -2)
\,E_{2}-M\,\Phi_{0}=0\,,
\eqno(1a)
$$

$$
i\,\epsilon\,E_{1}- \gamma  (a -ib) \,e^{z}\,H_{2}+
 i\, ({d \over dz}- 1) \,H_{1}-M\,\Phi_{1}=0\,,
$$
$$
i\,\epsilon\,E_{2}-  \gamma (a +ib) \,e^{z}\,H_{1} - \gamma(a -
ib)  \,e^{z}\, H_{3} -M\,\Phi_{2}=0\,,
$$
$$
i\,\epsilon\,E_{3}- \gamma ( a +ib) \,e^{z}\,H_{2} -i (\,{d \over
dz}-1) \,H_{3}-M\,\Phi_{3}=0\,, \eqno(1b)
$$

$$
-i\,\epsilon\,\Phi_{1}+ \gamma\;  (b+ ia) \,e^{z}\,\Phi_{0}
-M\,E_{1}=0\,,
$$
$$
-i\,\epsilon\,\Phi_{2} - {d \Phi_{0}\over dz}-M\,E_{2}=0\,,
$$
$$
-i\,\epsilon\,\Phi_{3}  + \gamma (b -ia)
\,e^{z}\,\Phi_{0}-M\,E_{3}=0\,, \eqno(1c)
$$

$$
\gamma (a -ib) \,e^{z}\,\Phi_{2}  - i ({d \over dz} -1)
\,\Phi_{1}-M\,H_{1}=0\,,
$$
$$
\gamma ( a+ib) \,e^{z}\, \Phi_{1} +  \gamma (a-ib) \,e^{z}\;
\Phi_{3}-M\,H_{2}=0\,,
$$
$$
\gamma (a+ib) \,e^{z}\,\Phi_{2}  + i\,({d \over dz}-1) \,\Phi_{3}
- M\,H_{3}=0\,. \eqno(1d)
$$

When performing nonrelativistic approximation  in (1) we will adhere
the method elaborated in \cite{2, 3}.

First,  with the help of
 $(1a)$ and  $(1d)$ let us exclude non-dynamical variables  $\Phi_{0}, H_{1},
H_{2}, H_{3}$  in eqs. $(1b),\;(1c)$  (components that were not differentiated in time)
$$
\Phi_{0}={1\over M}\,\left[\gamma (i a  - b ) e^{z}\,E_{1} -
\gamma (ia +b) e^{z} E_{3} - ({d \over dz} -2) \,E_{2}\right] ,
$$
$$
H_{1}={1\over M}\,\left[\gamma (a -ib) \,e^{z}\,\Phi_{2}  - i ({d
\over dz} -1) \,\Phi_{1}\right],
$$
$$
H_{2}={1\over M}\,\left[\gamma ( a+ib) \,e^{z}\, \Phi_{1} + \gamma
(a-ib) \,e^{z}\; \Phi_{3}\right] ,
$$
$$
H_{3}={1\over M}\,\left[\gamma (a+ib) \,e^{z}\,\Phi_{2}  + i\,({d
\over dz}-1) \,\Phi_{3}\right] ;
$$

\noindent  which results in
$$
i\,\epsilon\,E_{1}- \gamma  (a -ib) \,e^{z}\,{1\over M}\,
\left[\gamma ( a+ib) \,e^{z}\, \Phi_{1} +  \gamma (a-ib) \,e^{z}\;
\Phi_{3}\right]+
$$
$$
+
 i\, ({d \over dz}- 1) \,{1\over M}\,\left[\gamma (a -ib) \,e^{z}\,\Phi_{2}  - i ({d \over dz} -1) \,\Phi_{1}\right]-M\,\Phi_{1}=0\,,
$$
$$
-i\epsilon\Phi_{1}+ \gamma  (b+ ia) e^{z}{1\over
M}\left[\gamma (i a  - b ) e^{z}E_{1} - \gamma (ia +b) e^{z}
E_{3} - ({d \over dz} -2) E_{2}\right] -ME_{1}=0\,,
$$
$$
i\,\epsilon\,E_{2}-  \gamma (a +ib) \,e^{z}\,{1\over
M}\,\left[\gamma (a -ib) \,e^{z}\,\Phi_{2}  -
 i ({d \over dz} -1) \,\Phi_{1}\right] -
 $$
 $$
 -
\gamma(a - ib)  \,e^{z}\, {1\over M}\,\left[\gamma (a+ib)
\,e^{z}\,\Phi_{2}  + i\,({d \over dz}-1) \,\Phi_{3}\right]
-M\,\Phi_{2}=0\,,
$$
$$
-i\epsilon\Phi_{2} - {d \over dz}{1\over M}\left[\gamma (i a
- b ) e^{z}E_{1} - \gamma (ia +b) e^{z} E_{3} - ({d \over dz}
-2) E_{2}\right]-ME_{2}=0\,,
$$
$$
i\epsilon E_{3}- \gamma ( a +ib) e^{z}{1\over
M}\left[\gamma ( a+ib) e^{z} \Phi_{1} +
 \gamma (a-ib) e^{z} \Phi_{3}\right] -
 $$
 $$
 - i ({d \over dz}-1) {1\over M}\left[\gamma (a+ib) e^{z}\Phi_{2}  +
i({d \over dz}-1) \Phi_{3}\right]-M\Phi_{3}=0\,,
$$
$$
-i\epsilon\Phi_{3}  + \gamma (b -ia) e^{z}{1\over
M}\left[\gamma (i a  - b ) e^{z}E_{1} - \gamma (ia +b) e^{z}
E_{3} - ({d \over dz} -2) E_{2}\right]-ME_{3}=0\,.
$$
$$
\eqno(2)
$$

\noindent  In the system (2),  let  us  introduce big and small components,  $\Psi_{j}$  and $\psi_{j}$:
$$
\Phi_{j} = \Psi_{j} + \psi_{j}\;, \qquad   E_{j} = -i(\Psi_{j} -
\psi_{j})\,,
$$

\noindent  eqs.  (2) take the form
$$
\epsilon\,(\Psi_{1} - \psi_{1})- \gamma  (a -ib) \,e^{z}\,{1\over
M}\,\left[\gamma ( a+ib) \,e^{z}\, (\Psi_{1} + \psi_{1}) +  \gamma
(a-ib) \,e^{z}\; (\Psi_{3} + \psi_{3})\right]+
$$
$$
+
 i ({d \over dz}- 1) {1\over M}\left[\gamma (a -ib) e^{z}(\Psi_{2} + \psi_{2})
  - i ({d \over dz} -1) (\Psi_{1} + \psi_{1})\right]-M(\Psi_{1} + \psi_{1})=0\,,
$$

$$
\epsilon\,(\Psi_{1} + \psi_{1})+ \gamma\;  (b+ ia)
\,e^{z}\,{1\over M}\, \left[\gamma (i a  - b ) e^{z}\,(\Psi_{1} -
\psi_{1}) - \gamma (ia +b) e^{z} (\Psi_{3} - \psi_{3}) - \right.
$$
$$
\left. - ({d \over dz} -2) \,(\Psi_{2} - \psi_{2})\right]
-M\,(\Psi_{1} - \psi_{1})=0\,,
$$

$$
\epsilon(\Psi_{2} - \psi_{2})-  \gamma (a +ib) e^{z}{1\over
M}\left[\gamma (a -ib) e^{z}(\Psi_{2} + \psi_{2})  - i ({d
\over dz} -1) (\Psi_{1} + \psi_{1})\right] -
$$
$$-
\gamma(a - ib)  e^{z} {1\over M}\,\left[\gamma (a+ib)
e^{z}(\Psi_{2} + \psi_{2})  + i({d \over dz}-1) (\Psi_{3}
+ \psi_{3})\right] -M(\Psi_{2} + \psi_{2})=0\,,
$$

$$
\epsilon\,(\Psi_{2} + \psi_{2}) - {d \over dz}{1\over
M}\,\left[\gamma (i a  - b ) e^{z}\,(\Psi_{1} - \psi_{1}) - \gamma
(ia +b) e^{z} (\Psi_{3} - \psi_{3}) -  \right.
$$
$$
\left. - ({d \over dz} -2) \,(\Psi_{2} -
\psi_{2})\right]-M\,(\Psi_{2} - \psi_{2})=0\,,
$$

$$
\epsilon\,(\Psi_{3} - \psi_{3})- \gamma ( a +ib) \,e^{z}\,{1\over
M}\,\left[\gamma ( a+ib) \,e^{z}\, (\Psi_{1} + \psi_{1}) +  \gamma
(a-ib) \,e^{z}\; (\Psi_{3} + \psi_{3})\right]-
$$
$$ -i ({d \over dz}-1) {1\over M}\left[\gamma (a+ib) e^{z}(\Psi_{2} + \psi_{2})  +
i({d \over dz}-1) (\Psi_{3} + \psi_{3})\right]-M(\Psi_{3} +
\psi_{3})=0\,,
$$

$$
\epsilon\,(\Psi_{3} + \psi_{3})  + \gamma (b -ia) \,e^{z}\,{1\over
M}\, \left[\gamma (i a  - b ) e^{z}\,(\Psi_{1} - \psi_{1}) -
\gamma (ia +b) e^{z} (\Psi_{3} - \psi_{3}) - \right.
$$
$$
\left. -
 ({d \over dz} -2) \,(\Psi_{2} - \psi_{2})\right]-M\,(\Psi_{3} - \psi_{3})=0\,.
$$
$$
\eqno(3)
$$

Within each pair of equations, let  us sum and subtract two equations

$$
2\epsilon\Psi_{1}-2M\Psi_{1}+{1\over M}\left({d^{2}\over
dz^{2}}-2{d\over
dz}+1\right)(\Psi_{1}+\psi_{1})-{2\gamma^{2}e^{2z}(a^{2}+b^{2})\over
M}\Psi_{1}-
$$$$
-{2i\gamma e^{z}(ib-a)\over
M}(\Psi_{2}-\psi_{2})-{2i\gamma e^{z} (ib-a)\over M} {d
\psi_{2}\over dz}-{2\gamma^{2} e^{2z}(ib-a)^{2}\over
M}\,\psi_{3}=0\,,
$$

\vspace{3mm}

$$
-2\epsilon\psi_{1}-2M\psi_{1}+{1\over M}\left({d^{2}\over
dz^{2}}-2{d\over
dz}+1\right)(\Psi_{1}+\psi_{1})-{2\gamma^{2}e^{2z}(a^{2}+b^{2})\over
M}\psi_{1}+
$$
$$
+{2i\gamma e^{z}(ib-a)\over
M}(\Psi_{2}-\psi_{2})-{2i\gamma e^{z}(ib-a)\over M}{d
\Psi_{2}\over dz}-{2 \gamma^{2}e^{2z} (ib-a)^{2}\over
M} \Psi_{3}=0\,,
$$

\vspace{3mm}

$$
2\,\epsilon \Psi_{2}-2 M \Psi_{2}-{2 i \gamma
e^{z}  (a+ib)\over M} \Psi_{1}+
$$
$$
+ {2 i \gamma e^{z}(a+ib)\over M} {d \psi_{1}\over
dz}-{2 \gamma^{2} e^{2z} (a^{2}+b^{2})\over
M} (\Psi_{2}+\psi_{2})+
$$
$$
+{1\over M} \left({d^{2}\over dz^{2}}-2 {d\over
dz}\right) (\Psi_{2}-\psi_{2})-{2 i\gamma e^{z} (-a+ib)\over
M} \Psi_{3} +{2 i\gamma e^{z} (-a+ib)\over M} {d
\psi_{3}\over dz}=0\,,
$$

\vspace{3mm}

$$
-2 \epsilon \psi_{2}-2  M \psi_{2}-{2 i \gamma
e^{z} (a+ib)\over M} \psi_{1}+
$$
$$
+ {2 i \gamma e^{z}(a+ib)\over M} {d \Psi_{1}\over
dz}-{2 \gamma^{2} e^{2z} (a^{2}+b^{2})\over
M} (\Psi_{2}+\psi_{2})-
$$
$$
-{1\over M} \left({d^{2}\over dz^{2}}-2 {d\over
dz}\right) (\Psi_{2}-\psi_{2})-{2  i\gamma e^{z} (-a+ib)\over
M} \psi_{3} +{2 i\gamma e^{z} (-a+ib)\over M} {d
\Psi_{3}\over dz}=0\,,
$$

\vspace{3mm}

$$
2 \epsilon \Psi_{3}-2 M \Psi_{3}-
{2 \gamma^{2} e^{2z} (a+ib)^{2}\over M} \psi_{1}-
$$
$$
- {2 i \gamma e^{z} (a+ib)\over M} {d \psi_{2}\over
dz}-{2 i \gamma e^{z} (a+ib)\over M} (\Psi_{2}-\psi_{2})+
$$
$$
+{1\over M} \left({d^{2}\over dz^{2}}-2 {d\over
dz}+1\right) (\Psi_{3}+\psi_{3})-{2 \gamma^{2} e^{2z} (a^{2}+b^{2})\over
M} \Psi_{3}=0\,,
$$

\vspace{3mm}

$$
-2 \epsilon \psi_{3}-2 M \psi_{3}-{2 \gamma^{2} e^{2z}
(a+ib)^{2}\over M} \Psi_{1}-
$$
$$
- {2 i \gamma e^{z} (a+ib)\over M} {d \Psi_{2}\over
dz}+{2 i \gamma e^{z} (a+ib)\over M} (\Psi_{2}-\psi_{2})+
$$
$$
+{1\over M} \left({d^{2}\over dz^{2}}-2 {d\over
dz}+1\right) (\Psi_{3}+\psi_{3})-{2 \gamma^{2} e^{2z} (a^{2}+b^{2})\over
M} \psi_{3}=0\,.
$$
$$
\eqno(4)
$$

\noindent Next step is to separate  the rest energy
$$
\Psi = e^{-i \epsilon t} = e^{-i(m +E)t}  \;, \qquad i{\partial
\over \partial t}  e^{-i \epsilon t} = (M + E) e^{-i \epsilon t}
\; ,
$$
to this end it is enough in eqs.  (4) to perform  one formal change   $\epsilon$  on $(M+E)$:
$$
2E\Psi_{1}+{1\over M}\left({d^{2}\over dz^{2}}-2{d\over
dz}+1\right)(\Psi_{1}+\psi_{1})-{2\gamma^{2}e^{2z}(a^{2}+b^{2})\over
M}\Psi_{1}-
$$$$
-{2i\gamma e^{z}(ib-a)\over
M}(\Psi_{2}-\psi_{2})-{2i\gamma e^{z}(ib-a)\over M}{d
\psi_{2}\over dz}-{2\gamma^{2}e^{2z}(ib-a)^{2}\over
M}\psi_{3}=0\,,
$$

\vspace{3mm}

$$
-2E\psi_{1}+{1\over M}\left({d^{2}\over dz^{2}}-2{d\over
dz}+1\right)(\Psi_{1}+\psi_{1})-{2\gamma^{2}e^{2z}(a^{2}+b^{2})\over
M}\psi_{1}+
$$$$
+{2i\gamma e^{z}(ib-a)\over
M}(\Psi_{2}-\psi_{2})-{2i\gamma e^{z}(ib-a)\over M}{d
\Psi_{2}\over dz}-{2\gamma^{2}e^{2z}(ib-a)^{2}\over
M}\Psi_{3}=4M\psi_{1}\,,
$$

\vspace{3mm}

$$
2E\Psi_{2}-{2i\gamma e^{z}(a+ib)\over
M}\Psi_{1}+{2i\gamma e^{z}(a+ib)\over M}{d \psi_{1}\over
dz}-{2\gamma^{2}e^{2z}(a^{2}+b^{2})\over
M}(\Psi_{2}+\psi_{2})+
$$
$$
+{1\over M}\left({d^{2}\over dz^{2}}-2{d\over
dz}\right)(\Psi_{2}-\psi_{2})-{2i\gamma e^{z}(-a+ib)\over
M}\Psi_{3} +{2i\gamma e^{z}(-a+ib)\over M}{d
\psi_{3}\over dz}=0\,,
$$

\vspace{3mm}

$$
-2E\psi_{2}-{2i\gamma  e^{z}(a+ib)\over
M}\psi_{1}+{2i\gamma e^{z}(a+ib)\over M}{d \Psi_{1}\over
dz}-{2\gamma^{2}e^{2z}(a^{2}+b^{2})\over
M}(\Psi_{2}+\psi_{2})-
$$
$$
-{1\over M}\left({d^{2}\over dz^{2}}-2{d\over
dz}\right)(\Psi_{2}-\psi_{2})-{2i\gamma e^{z}(-a+ib)\over
M}\psi_{3} +{2i\gamma e^{z}(-a+ib)\over M}{d
\Psi_{3}\over dz}=4 M\psi_{2}\,,
$$

\vspace{3mm}

$$
2E\Psi_{3}-{2\gamma^{2}e^{2z}(a+ib)^{2}\over
M}\psi_{1}-{2i\gamma e^{z}(a+ib)\over M}{d
\psi_{2}\over dz}-{2i\gamma e^{z}(a+ib)\over
M}(\Psi_{2}-\psi_{2})+
$$
$$
+{1\over M}\left({d^{2}\over dz^{2}}-2{d\over
dz}+1\right)(\Psi_{3}+\psi_{3})-{2\gamma^{2}e^{2z}(a^{2}+b^{2})\over
M}\Psi_{3}=0\,,
$$

\vspace{3mm}

$$
-2E\psi_{3}-{2\gamma^{2}e^{2z}(a+ib)^{2}\over
M}\Psi_{1}-{2i\gamma e^{z}(a+ib)\over M}{d
\Psi_{2}\over dz}+{2i\gamma e^{z}(a+ib)\over
M}(\Psi_{2}-\psi_{2})+
$$
$$
+{1\over M}\left({d^{2}\over dz^{2}}-2{d\over
dz}+1\right)(\Psi_{3}+\psi_{3})-{2\gamma^{2}e^{2z}(a^{2}+b^{2})\over
M}\psi_{3}=4M\psi_{3}\,.
$$
$$\eqno(5)
$$

Conditions of non-relativity reduce to two restrictions:
$$
E << M \; , \qquad \psi_{j} << \Psi_{j}\,.\eqno(6)
$$

\noindent Taking into account (6), from  $(5)$ one can derive three equations for big components
 $\Psi_{j}$   (also there arise  expressions for small components in terms of big ones).
 Below we  write down only three equations for big components
 $$
\left(  {d^{2}\over dz^{2}}-2\,{d\over dz}+ 2\,E\,M  + 1 -
 e^{2z}\,(a^{2}+b ^{2}) \right)\,\Psi_{1}
+2\,i\,\gamma\,e^{z} \,(a - ib) \,\Psi_{2}=0\,, \eqno(7a)
$$

$$
\left (
 {d^{2}\over dz^{2}}-2\,{d\over dz}+ 2\,E\,M  +1  -
e^{2z}\,(a^{2}+b^{2}) \right ) \Psi_{3} -
2\,i\,\gamma\,e^{z}\,(a+ib)\,\Psi_{2} =0\,, \eqno(7b)
$$

$$
 \left( {d^{2}\over dz^{2}}-2\,{d\over dz}           +  2\,E\,M\,-
e^{2z}\,(a^{2}+b^{2})  \right)\,\Psi_{2} -
$$
$$
- 2\,i\,\gamma \,e^{z}\,(a+ib) \,\Psi_{1}  +
2\,i\gamma\,e^{z}\,(a-ib)\,\Psi_{3} =0\,. \eqno(7c)
$$

\noindent
  These are what we need: the system of equations in Pauli approximation.

Let us consider how these equations behave when vanishing curvature.
To this end, it is better to translate   (7)  to usual units
$$
E = {\epsilon \rho \over \hbar c}\;, \qquad M = {mc\rho \over
\hbar } \;, \qquad 2EM= 2m\epsilon \;{\rho^{2} \over \hbar^{2}}\;,
$$
$$
a = {P_{1} \rho \over \hbar}\;, \qquad  b = {P_{2} \rho \over
\hbar}\;, \qquad z = {Z \over \rho }\; .
$$

For instance, the first equation takes the form
$$
{2\,\epsilon \,m \over \hbar^{2} }
 \,\Psi_{1} + \left({d^{2}\over dZ^{2} }-{2\over \rho}  \,{d\over dZ}+ {1 \over \rho^{2}} \right)\,\Psi_{1}-
$$
$$
-  \,e^{2Z /\rho} \,{P_{1}^{2}+P_{2}^{2} \over \hbar^{2} }
\,\Psi_{1} - 2\,i\,\gamma\,e^{Z/\rho} \,{ i P_{2}-P_{1} \over
\hbar \rho} \; \Psi_{2}=0\,;
$$

\noindent at vanishing curvature,  $\rho \rightarrow \infty $, it reduces to
$$
{2\,\epsilon \,m \over \hbar^{2} }
 \,\Psi_{1} + {d^{2}\over dZ^{2} }\,\Psi_{1}-  {P_{1}^{2}+P_{2}^{2} \over \hbar^{2} } \,\Psi_{1} =0\,.
\eqno(8a)
$$

Similarly,  one can consider two remaining equations -- so we get
$$
{2\,\epsilon \,m \over \hbar^{2} }
 \,\Psi_{1} + {d^{2}\over dZ^{2} }\,\Psi_{1}-  {P_{1}^{2}+P_{2}^{2} \over \hbar^{2} } \,\Psi_{2} =0\,,
\eqno(8b)
$$
$$
{2\,\epsilon \,m \over \hbar^{2} }
 \,\Psi_{1} + {d^{2}\over dZ^{2} }\,\Psi_{1}-  {P_{1}^{2}+P_{2}^{2} \over \hbar^{2} } \,\Psi_{3} =0\,.
\eqno(8c)
$$

Equations  (8) coincide with those arising when examining the same problem in Minkowski space
$$
( 2\,EM - a^{2}-b^{2} - p_{3}^{2} ) \;\Psi_{1}=0\,,
$$
$$
( 2\,EM - a^{2}-b^{2} - p_{3}^{2} ) \;\Psi_{2}=0\,,
$$
$$
( 2\,EM - a^{2}-b^{2} - p_{3}^{2} ) \;\Psi_{3}=0\, . \eqno(9)
$$

Note that the most simple solutions for equations (9) in Minkowski space taken  according to
$$
\Psi_{1}\neq 0\,, \qquad \Psi_{2}=0\,,\qquad \Psi_{3}=0\,,
$$
$$
\Psi_{1}=0\,, \qquad \Psi_{2} \neq 0\,,\qquad \Psi_{3}=0\,,
$$
$$
\Psi_{1}=0\,, \qquad \Psi_{2}=0\,,\qquad \Psi_{3} \neq 0\,,
$$

\noindent do not provide us with eigenstates of helicity operator for spin 1 field.

To construct solutions for system  (7) in Lobachevsky space, below we will use
 additional operator, generalized helicity operator  $\Sigma$.
 Definition and main properties of that were given in \cite{1}.
 In nonrelativistic limit, the eigenvalue equation
  $\Sigma \Psi =
\sigma \Psi$  leads to tree relations
$$
\gamma \,(a-ib) \,e^{z}\,\Psi_{2} = \left ( + i\,({d \over dz}-1)
+\sigma\, \right ) \Psi _{1}\,,
$$
$$
 \gamma (a+ib) \,e^{z}\,\Psi_{1}+
\gamma\,(a-i b) \,e^{z}\,\Psi _{3}=\sigma\,\Psi_{2}\,,
$$
$$
\gamma (a+ib) \,e^{z}\,\Psi _{2} = \left ( - i\,({d \over dz}-1) +
\sigma\, \right ) \Psi_{3}\,. \eqno(10)
$$

\noindent
Depending on values of $\sigma$ (zero or non-zero) we have substantially different
constrains in (10).

First, let us examine the case $\sigma \neq 0$ (from general consideration it is evident that
in the nonrelativistic limit the following relation must hold: $\sigma =   \pm \sqrt{2ME} $)
$$
\gamma \,(a-ib) \,e^{z}\,\Psi_{2} = \left ( + i\,({d \over dz}-1)
+\sigma\, \right ) \Psi _{1}\,,
$$
$$
 \gamma (a+ib) \,e^{z}\,\Psi_{1}+
\gamma\,(a-i b) \,e^{z}\,\Psi _{3}=\sigma\,\Psi_{2}\,,
$$
$$
\gamma (a+ib) \,e^{z}\,\Psi _{2} = \left ( - i\,({d \over dz}-1) +
\sigma\, \right ) \Psi_{3}\,.
 \eqno(11a)
$$

\noindent Because  $\sigma \neq 0$, with the help of second relation  one can exclude the  function
 $\Psi_{2}$
$$
\Psi_{2} =    {\gamma \over \sigma }  e^{z}  \left [  (a+ib)
\,\Psi_{1}+ (a-i b) \, \Psi _{3} \right ]
 \eqno(11b)
 $$

\noindent
  in the first and the third equations, thus  obtaining the system for only two functions
  $$
(a-i b)^{2} e^{2z} \Psi _{3}
 = \left [  + 2i \sigma ({d \over dz}-1)
+2\sigma^{2}    -  (a^{2} +b^{2})  e^{2z}  \right ] \Psi_{1} \,,
$$
$$
 (a+ib)^{2} e^{2z}  \Psi_{1}
 = \left [   - 2i\sigma ({d \over dz}-1)
+ 2\sigma^{2} \,   - (a^{2} + b^{2} ) e^{2z} \right ] \Psi_{3}\,.
 \eqno(11c)
 $$

Let us find an  a second order equation for $\Psi_{1}$:
$$
 (a+ib)^{2} e^{2z}  \Psi_{1}
 = \left [   - 2i\sigma ({d \over dz}-1)
+ 2\sigma^{2} \,   - (a^{2} + b^{2} ) e^{2z} \right ] \times
$$
$$
\times {e^{-2z}  \over (a-i b)^{2} }
 \left [  + 2i \sigma ({d \over dz}-1)
+2\sigma^{2}    -  (a^{2} +b^{2})  e^{2z}  \right ] \Psi_{1} \,,
 $$

\noindent or
$$
 (a^{2} +b^{2}) ^{2} e^{4z}  \Psi_{1}
 =
 $$
 $$
 =
 \left [   - 2i\sigma ({d \over dz}-3)
+ 2\sigma^{2} \,   - (a^{2} + b^{2} ) e^{2z} \right ] \times
$$
$$
\times
 \left [  + 2i \sigma ({d \over dz}-1)
+2\sigma^{2}    -  (a^{2} +b^{2})  e^{2z}  \right ] \Psi_{1} \,,
 $$

\noindent and further
$$
 (a^{2} +b^{2}) ^{2} e^{4z}  \Psi_{1}
 =
 $$
 $$
 =\left [\;
 4\sigma^{2}  ({d \over dz}-3) ({d \over dz}-1) - i\sigma 4\sigma^{2} ({d \over dz}-3)
 + 2i\sigma (a^{2} + b^{2}) e^{2z}  ({d \over dz}-1) + \right.
$$
$$
+ \sigma^{2} 4i \sigma ({d \over dz}-1) + 4\sigma^{2} \sigma^{2}
-2\sigma^{2} (a^{2} +b^{2})  e^{2z}-
$$
$$
\left. - (a^{2} + b^{2} ) e^{2z} 2i \sigma ({d \over dz}-1)  -
(a^{2} + b^{2} ) e^{2z}  2\sigma^{2} + (a^{2} +b^{2})^{2}  e^{4z}
\right  ] \Psi_{1}\;.
$$

\noindent From whence, after evident simplifying awe arrive at
$$
0  = 4\sigma^{2} \left [\;
   ({d^{2} \over dz^{2}} -4 {d\over dz} +3)  + 2 i\sigma
 +  \sigma^{2} -(a^{2} +b^{2})  e^{2z} \right  ] \Psi_{1}
$$

\noindent  that is
$$
\left [\;
   {d^{2} \over dz^{2}} -4 {d\over dz} +  \sigma^{2} +3  + 2 i\sigma
  - (a^{2} +b^{2})  e^{2z} \right  ] \Psi_{1} = 0\; .
\eqno(12a)
$$

\noindent Similarly, equation for $\Psi
_{3}$ is
$$
\left [  {d^{2}\over dz^{2}}-4 \,{d\over dz}  + \sigma^{2}  +3
-2i\sigma  - e^{2z}\,(a^{2}+b ^{2})
  \right ] \Psi _{3}=0\,.
\eqno(12b)
$$

Remembering, that  $\Psi _{1}, \; \Psi _{3}$ $(12a,b)$ are not independent, instead they are connected
by the first order differential system  $(11c)$.  Moreover, relation  $(11b)$
permits us to construct  explicitly  $\Psi _{2}$ in terms of the known
$\Psi _{1}, \; \Psi _{3}$. which provides us with the complete solving of the problem.

In order to detail the problem, let us translate eqs.  $(12a)$ and  $(12b)$ to a new variable
$$
x=i\,\sqrt{a^{2}+b^{2}}\;e^{z}\,,
$$
$$
\left[x\,{d^{2}\over dx^{2}}-3\,{d\over dx}+x+{(\sigma-i)\,(\sigma+3i)\over x}\right]\Psi_{1}=0\,,
$$
$$
\left[x\,{d^{2}\over dx^{2}}-3\,{d\over
dx}+x+{(\sigma+i)\,(\sigma-3i)\over x}\right]\Psi_{3}=0\,.
\eqno(13a)
$$

Through  the substitutions
$$
\Psi_{1} = x^{2} f_{1} (x)\;, \qquad \Psi_{3} = x^{2} f_{3} (x)\;,
$$

\noindent
eqs. $(13a)$ reduce to the Bessel form
$$
{d^{2} \over dx^{2}} f_{1} + {1 \over x} {d \over dx} f_{1} +
\left ( 1 -  {(1- i \sigma )^{2} \over x^{2}} \right ) f_{1} = 0\,,
$$
$$
{d^{2} \over dx^{2}} f_{3} + {1 \over x} {d \over dx} f_{3} +
\left ( 1 -  {(1+ i \sigma )^{2} \over x^{2}} \right ) f_{3} = 0
\eqno(13b)
$$

\noindent
with the following solutions
 (let it be:    $\nu = 1 - i \sigma, \; \mu = 1 + i \sigma $)
$$
f^{+} _{1}(x) = A^{+}_{1} J_{ + \nu}(x)\;, \qquad f^{-} _{1}(x) =  A^{-}_{1} J_{ - \nu}(x)\;,
$$
$$
f^{+} _{3}(x) =  A^{+}_{3} J_{ + \mu }(x)\;, \qquad f^{-} _{3}(x) =  A^{-}_{3} J_{ -\mu }(x)\; .
\eqno(13c)
$$

We wish to know  pairs of solutions in $(13c)$ that are linked by
the first order relations $(11c)$.
To clarify that let us translate eqs.  $(11c)$ to the  above variable $x$:
$$
x=i\,\sqrt{a^{2}+b^{2}}\;e^{z}\,,
$$
$$
\left ( +2\,i\,\sigma\,x\,{d  \over dx}-2\,i\,\sigma\, +2\,\sigma^{2} +x^{2} \right ) \Psi_{1}  +
{(a-i\,b) \over a + i b }\; x^{2}  \Psi_{3}=0\,,
$$
$$
\left ( -2\,i\,\sigma\,x\,{d \over dx}+2\,i\,\sigma +2\,\sigma^{2} +x^{2} \right ) \Psi_{3} +
 {(a+i\,b) \over a - ib  }\; x^{2} \Psi_{1}=0\,,
$$
$$
\eqno(14a)
$$

\noindent by substitutions  they reduce to
$$
\Psi_{1} = x^{2} f_{1} (x)\;, \qquad \Psi_{3} = x^{2} f_{3} (x)\;,
$$
$$
\left ( +2\,i\,\sigma\,x\,{d  \over dx}+2\,i\,\sigma\, +2\,\sigma^{2} +x^{2} \right ) f_{1}  +
{(a-i\,b) \over a + i b }\; x^{2}  f_{3}=0\,,
$$
$$
\left ( -2\,i\,\sigma\,x\,{d \over dx}-2\,i\,\sigma +2\,\sigma^{2} +x^{2} \right ) f_{3} +
 {(a+i\,b) \over a - ib  }\; x^{2} f_{1}=0\,.
$$
$$
\eqno(14b)
$$

\noindent It is easily to simplify the formulas by the following notation
$$
(a-ib)f_{3} = \bar{f}_{3}, \qquad  (a+ib)f_{1} = \bar{f}_{1}\; ,
$$
$$
\left ( +2\,i\,\sigma\, (x\,{d  \over dx} + \nu)  +x^{2} \right ) \bar{f}_{1}  +
 x^{2}  \bar{f}_{3}=0\,,
$$
$$
\left ( -2\,i\,\sigma\, (x\,{d \over dx} + \mu)   +x^{2} \right ) \bar{f}_{3} +
 x^{2} \bar{f}_{1}=0\,.
\eqno(14с)
$$

Let us demonstrate that one possible solution
is realized on the functions
$$
\bar{f}_{1} = J_{ + \nu} (x) , \qquad  \bar{f}_{3} = J_{ - \mu} (x) \;,
$$
$$
 + \nu = 1 - i \sigma,  \qquad- \mu  = -1 - i \sigma  \; .
\eqno(15a)
$$

\noindent
To this end, it it suffices to
 apply the first equation
 $$
\left ( +2\,i\,\sigma\, (x\,{d  \over dx} + \nu)  +x^{2} \right ) J_{\nu} (x)  +
 x^{2}  J_{-\mu}(x) =0\,.
$$

\noindent
With the help of the known relation for Bessel functions
$$
(x {d \over dx}  + \nu ) J_{\nu} = + x J_{\nu -1}
$$

\noindent  it can be translated to the form
$$
 2 i \sigma \;   J_{-i \sigma } (x)    + \left [\; x  J_{-i\sigma +1  } (x) +
 x  J_{-i\sigma -1 } (x) \; \right ] =0\,.
\eqno(15b)
$$

\noindent
From whence, allowing for the known recursive relation
$$
x J_{a+1}  + x J_{a-1} =  2 a J_{a} \; , \qquad \mbox{when} \qquad
 a=-i\sigma \; ,
$$

\noindent
we get an identity $0=0$.

Thus, we have constructed the firs class of  solutions

\vspace{3mm}
$\sigma = \pm \sqrt{2ME}$,
$$
\Psi^{I}_{1} = {x^{2}   \over a+ib} \; J_{+\nu}(x)\;, \qquad  \Psi_{3}^{I} = {x^{2}  \over a-ib} \;  J_{-\mu}(x)\; .
$$
$$
\Psi_{2}^{I} =    {\gamma \over \sigma }  e^{z}  \left [  (a+ib)
\,\Psi_{1}^{I}+ (a-i b) \, \Psi _{3} ^{I} \right ]=
$$
$$
=
{\gamma \over \sigma }  e^{z}  \left [
 x^{2}  \; J_{+\nu}(x) +  x^{2}   \;  J_{-\mu}(x)  \right ]\,.
\eqno(15c)
$$

Similarly, we can prove existent of the second class of  solutions

\vspace{3mm}
$\sigma = \pm \sqrt{2ME}$,
$$
\Psi_{1}^{II} = {x^{2}   \over a+ib} \; J_{-\nu}(x)\;, \qquad  \Psi_{3}^{II}  = {x^{2}  \over a-ib} \;  J_{+\mu}(x)\; .
$$
$$
\Psi_{2}^{II} =    {\gamma \over \sigma }  e^{z}  \left [  (a+ib)
\,\Psi_{1}^{II}+ (a-i b) \, \Psi _{3} ^{II} \right ]=
$$
$$
=
{\gamma \over \sigma }  e^{z}  \left [
 x^{2}  \; J_{-\nu}(x) +  x^{2}   \;  J_{+\mu}(x)  \right ]\,.
\eqno(15d)
$$

Now let us turn to the case of  $\sigma =0$,  it is specified by the following relations (see  (1))
$$
\Psi_{3} = - {a+ib \over a - ib} \; \Psi_{1}\;, \qquad \Psi_{2} =
+ {i e^{-z} \over \gamma (a -ib)} \,({d \over dz}-1)
 \Psi _{1}\, .
 \eqno(16)
$$

Allowing for   (13)  in  (7), we obtain
$$
\left (  {d^{2}\over dz^{2}}-4\,{d\over dz} + 2\,E\,M  +3   -
 e^{2z}\,(a^{2}+b ^{2}) \right
)  \Psi_{1}  =0\,, \eqno(17a)
$$
$$
\left (
 {d^{2}\over dz^{2}} -4\,{d\over dz} + 2\,E\,M\, +3   - e^{2z}\,(a^{2}+b^{2})  \right ) \Psi_{3} =0\,,
\eqno(17b)
$$

$$
\left({d^{2}\over dz^{2}}-2\,{d\over dz}  + 2\,E\,M  -
e^{2z}\,(a^{2}+b^{2}) \right) \Psi _{2}  +
$$
$$
+ 2\,i\gamma\,e^{z}\,(a-ib)\,\Psi_{3}  - 2\,i\,\gamma \,e^{z} \,
{(a+i b)} \Psi_{1} =0\,. \eqno(17c)
$$

The second relation in (16) says that  $\Psi_{2}$ can be found on the base of the known
 $\Psi_{1}$. In fact, this gives the complete solving of the problem when $\sigma =0$.

Note that equations for  $\Psi_{1}$ and $\Psi_{3}$ are the same. Let us translate
eq. $(17a, b)$  to the new variable
$$
x=i\,\sqrt{a^{2}+b^{2}}\;e^{z}\,,
$$
$$
\left (x\,{d^{2}\over dx^{2}}-3\,{d\over dx}+x+{3+2\,E\,M\over x}\right )\Psi_{1}=0\,,
$$
$$
\left ( x\,{d^{2}\over dx^{2}}-3\,{d\over
dx}+x+{3+2\,E\,M\over x}\right )\Psi_{3}=0\,.
\eqno(18a)
$$

By substitution $
\Psi_{1} = x^{2}\, g_{1} \,(x)$ it reduces to  the the Bessel type
$$
{d^{2} \over dx^{2}}\, g_{1} + {1 \over x}\, {d \over dx}\, g_{1} +
\left ( 1 -  {1-2\,E\,M \over x^{2}} \right ) g_{1} = 0\,.
\eqno(18b)
$$

Thus, at $\sigma =0$  there are constructed solutions of two types:

\vspace{3mm}
$\sigma =0$,
$$
\Psi^{I} _{1}(x) = +{x^{2} \over a+ib } J_{ + \nu}(x)\;, \qquad \Psi^{I} _{3}(x) = - {x^{2} \over a -ib}  J_{ +  \nu}(x)\;,
$$
$$
\Psi^{II} _{1}(x)  = + {x^{2} \over a+ib } J_{ - \nu}(x)\;, \qquad \Psi^{II} _{1}(x)  = - {x^{2} \over a -ib}  J_{ -  \nu}(x)\;.
$$
$$
\eqno(18c)
$$

It remains to prove that eq. $(17c)$ is consistent with (16) and  $(17a, b)$.
To this end, it is convenient to introduce new  functions
$$
(a+ib) \Psi_{1} = e^{z} \bar{\Psi}_{1}\;, \qquad
(a-ib) \Psi_{3} = e^{z} \bar{\Psi}_{3}\;,
\eqno(19a)
$$

\noindent
then
$$
\bar{\Psi}_{3} = -  \bar{\Psi}_{1}\;, \qquad
\Psi_{2} =  + {i  \over \gamma (a^{2} + b^{2})} \, {d \over dz}
 \bar{\Psi} _{1}\, .
\eqno(19b)
$$
$$
\left (  {d^{2}\over dz^{2}}-2\,{d\over dz} + 2\,E\,M    -
 e^{2z}\,(a^{2}+b ^{2}) \right
)  \bar{\Psi}_{1}  =0\,,
\eqno(19c)
$$
$$
\left({d^{2}\over dz^{2}}-2\,{d\over dz}  + 2\,E\,M  -
e^{2z}\,(a^{2}+b^{2}) \right) \Psi _{2}
 - 4\,i\,\gamma \,e^{2z} \,
\bar{\Psi}_{1} =0\,.
$$
$$
\eqno(19d)
$$

\noindent
Let us substitute  the expression for  $\bar{\Psi}_{2}$  $(19b)$ into   eq. $(19d)$:
$$
\left({d^{2}\over dz^{2}}-2\,{d\over dz}  + 2\,E\,M  -
e^{2z}\,(a^{2}+b^{2}) \right)  {d \over dz}
 \bar{\Psi} _{1}
 - 2 (a^{2} + b^{2}) e^{2z} \,
\bar{\Psi}_{1} =0\, .
$$

\noindent
Allowing for the following relation
$$
\left ( {d^{2} \over dz^{2} } - 2{d \over dz} + 2EM  \right ) {d \over dz}   \bar{\Psi}_{1} =
{d \over dz} e^{2z} (a^{2} + b^{2} ) \bar{\Psi}_{1} \; ;
$$

\noindent
we translate it to the form
$$
{d \over dz} e^{2z} (a^{2} + b^{2} ) \bar{\Psi}_{1}-
e^{2z}\,(a^{2}+b^{2})   {d \over dz}
 \bar{\Psi} _{1}
  - 2 (a^{2} + b^{2}) e^{2z} \,
\bar{\Psi}_{1} =0\, ,
$$

\noindent
which is an  identity.

Thus, the problem of solving the Duffin--Kemmer equation for spin 1 particle
on the background of  Lobachevsky space in Pauli approximation is completed.
A basis based on the four operators
$$
i {\partial \over  \partial t}, \;  - i {\partial \over \partial y}, \;  - i {\partial \over \partial x}, \;
\Sigma
$$

\noindent is constructed. Ultimately, the problem ends with a second order differential equation
of Schr\"{o}dinger type with a barrier of special form that generates a reflection coefficient $R=1$ for all states
-- for more detail see in \cite{5}.

Authors are grateful to V.M. Red'kov for encouragement  and advices.

\end{document}